# A topological gap waveguide based on unidirectional locking of pseudo-spins

Yan Ren[1], Hai Lin[1,a], Rui Zhou[1], Xintong Shi[1], Jing Jin[1], Y. Liu (刘泱杰)[2,b]

[1] College of Physics Science and Technology, Central China Normal University, Wuhan 430079;
renyan@mails.ccnu.edu.cn (Y.R.); ruizhou@mails.ccnu.edu.cn (R.Z.); shxt@mails.ccnu.edu.cn (X.S.); jingjin@ccnu.edu.cn (J.J.);

[2] Department of Physics, School of Physics, Hubei University, Wuhan 430062

[a] Corresponding author: linhai@mail.ccnu.edu.cn; Tel.: +86-186-2788-4302

[b] Corresponding author: yangjie@hubu.edu.cn; http://phys.hubu.edu.cn/info/1020/1078.htm

**Abstract**

Photonic topological insulators (PTIs) have been widely studied due to the robustness of energy transport via supported edge modes immune to structural disorder. In this work, a topological gap waveguide is constructed by introducing line defect into a topological photonic crystal structure and combining it with a gap waveguide structure, which design therefore combines the advantages of both topological and gap waveguides. Not only does it give high transmission efficiency, but also enables high robustness for energy transmission under structural defects and sharp bends. Our proposed topological waveguide design can be implemented with conventional semiconductor technology and integrated into optical circuits for communication systems.

## 1 introduction

With the discovery of the quantum Hall effect (QH) in condensed matter physics, the concept of topology came into being [1-3]. Strong spin-orbit coupling, such as the quantum spin Hall effect (QSH), can produce a class of exotic materials called topological insulators that are insulated in the bulk but conduct electricity along the surface through a class of topological edge states [4,5]. In particular, many photonic topological insulators (PTIs) have been proposed and shown to be analogues of topological insulators, where the degree of freedom of the electron spin can be simulated by constructing photonic pseudospins [6-8]. To design spin Hall topological photonic crystals, one takes advantage of electromagnetic (EM) duality [9,10] or utilizes crystal symmetry [11-13]. Interfaces between topological insulators with different topologies support strongly constrained topology-protected edge state [14,15]. Energy transport in these edge states is robust to defects and disorder that do not change the system topology, which has potential applications in improving the transmission quality and efficiency of optical signals [16-19]. The PTIs provide an excellent platform for the development of next-generation on-chip photonic devices.

Gap waveguide is a type of bed-of-nails contact-less structure with great advantages in broadband electromagnetic shielding characteristics and flexibility for building microwave structures. Its unique contact-less structure and broadband electromagnetic shielding characteristics have great advantages and flexibility in building new electromagnetic transmission line and shielding structure [20-22]. Compared with conventional waveguides, it is easier to process and more suitable for relatively high frequency and power capacity density [23-25]. Therefore, in the microwave band and above, the structure is suitable for waveguides, filters and other microwave passive and active devices. The existing studies on gap waveguides do not take full consideration of the topological characteristics of the structure. The introduction of non-trivial topology into the gap waveguide structure will open some new research perspectives and implementations in the field of microwave/millimeter wave devices, circuits and antennas [26].

Another useful feature to dynamically reconfigure on-demand of topological photonic crystals (PCs) has been crucial in recent practical applications. Various methods have been used to achieve tunable topological states, including mechanical control [10,27], thermally induced phase transitions in phase change materials [26,29], electrical tuning of graphene [30,31], and nematic liquid crystal (LC) rearrangement [32-34]. Among them, nematic LC are the most economical solution to realize tunable devices, which are subject to deformation as the LC molecules of the whole system rearrange under the action of external forces [35-37]. Because of this property, nematic LC is widely used in display technology, optical tunable devices, electronic components, and other applications.

In this work, we propose a topological gap waveguide based on spin-unidirectional locking with nematic LC. In it a gap sandwiches between nontrivial PTIs and meanwhile metal plates, which introduces topological line defects into the gap waveguide structure. Similar to topological edge states formed by trivial and nontrivial PCs, such defect mode states are immune to backscattering, unidirectional locking of energy transport, and robust to sharp bends and defects. This waveguide device combines the advantages of gap waveguide devices in fabrication process and the robustness characteristics of topological waveguide in energy transfer. The nematic LC is used as the background material of the structure, and an external circuit is applied to induce the rearrangement of the LC molecules, thus changing the physical properties of the background LC and realizing a tunable topological gap waveguide. The proposed waveguide device has potentials in achieving CMOS compatibility and backscattering suppression in tunable on-chip integrated topological photonic devices.

## 2 Structures and Methods

As shown in Fig. 1(a), we design a topological gap waveguide based on a nontrivial topological PC sandwiched between the top and bottom metal plates. The nontrivial topological PC consists of a triangular lattice of artificial molecules formed by six metallic cylinders with height ($h_0$) of 1mm. The height of the gap between the top plate and the PC structure is 0.25 mm. Geometry of the 2D PC arranged in a triangular lattice is demonstrated in inset of Fig. 1(a), where $a_0$ is the lattice constant, $\vec{a_1}$ and $\vec{a_2}$ are two translation vectors with the length of $a_0$. Black dashed rhombus and hexagon are primitive cells of honeycomb and triangular lattices, R is the distance from the center of the cylinder to the hexagon, and $d$ is the diameter of the cylinder. Therefore, our structure of the gap waveguide forms a so-called bed of nails topological on the bottom metallic plate, which is used as the bottom and side walls of the waveguide. The other plate is unpatterned and forms the top wall of the waveguide [20].

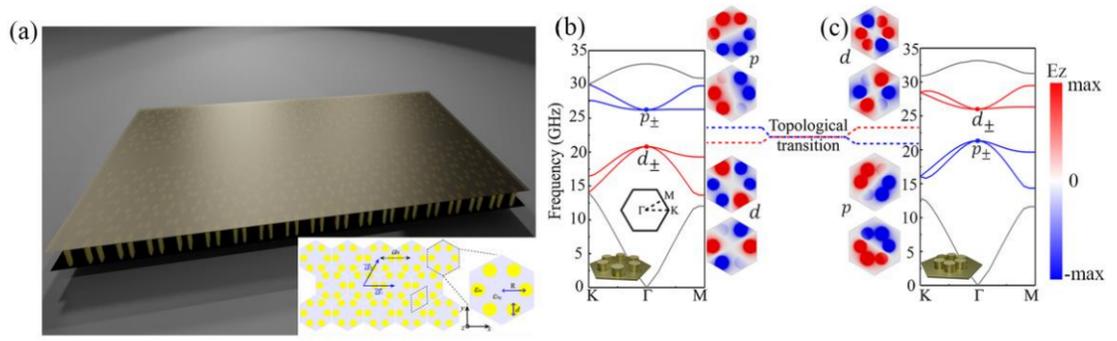

**Fig.1** (a) Schematic diagram of the topological gap waveguide structure, which is composed of a triangular lattice array. The bottom-right corner inset describes the geometry of the 2D PC arranged in a triangular lattice. $a_0$=9 mm is the lattice constant, $\vec{a_1}$ and $\vec{a_2}$ are two translation vectors with the length of $a_0$. Black dashed rhombus and hexagon are primitive cells of honeycomb and triangular lattices, R is the distance from the center of the cylinder to the hexagon, $d$=1 mm is the diameter of the cylinder. Corresponding dispersion relations of the topological (b) nontrivial structure, R=$a_0$/2.7, the hexagonal inset is the Brillouin zone of triangular lattice and bottom-left corner inset is the unit cell of nontrivial structure, (c) trivial structure, R=$a_0$/3.52,} bottom-left corner inset is the unit cell of trivial structure. The middle inset represents the topological modes inversion underlying the transition between pseudospin states.

Here, we consider the transverse magnetic (TM) mode with out-of-plane electric field and in-plane magnetic field, which is different from the transverse magnetic (TE) modes discussed in silicon-based valley photonic crystals [38-41]. For simplicity, the real permittivities of both cylinders ($\varepsilon_m$) and environment ($\varepsilon_{bg}$) are taken frequency independent in the regime under consideration. The master equation for a harmonic mode of frequency $\omega$ is derived from Maxwell equations [1,39]

$$\frac{1}{\varepsilon(\mathbf{r})}[\nabla \times \nabla \times E_z(\mathbf{r})\hat{z}] = \frac{\omega^2}{c^2}E_z(\mathbf{r})\hat{z} \, , \quad (1)$$

where $\varepsilon(\mathbf{r})$ represents the position-dependent permittivity and $c$ the speed of light. The Bloch theorem applies when $\varepsilon(\mathbf{r})$ is periodic, as shown in Fig.1. For our proposed PC structure, simply by shrinking and expanding the distance R while keeping the hexagonal clusters composed of six neighboring cylinders and the $C_6$ symmetry as seen in the bottom-left corner inset of Figs. 1(b-c). The metamolecules is reconfigured to generate the topological transition between trivial and nontrivial via pseudospin dipoles and quadrupoles. According to the characteristics of the $C_6$ symmetric, there are two 2D irreducible representation $E_1$ and $E_2$ [13], which correspond to the double degenerate points of the frequency bands, respectively. The irreducible representation $E_1$ corresponds to the double degenerate dipole state, i.e., two $p$ orbits: $p_x$ and $p_y$, and the irreducible representation $E_2$ corresponds to the double degenerate quadrupole state, i.e., two $d$ orbits: $d_{x^2-y^2}$ and $d_{xy}$. With dependence to the length R, Figs. 1(b-c) show the band inversion process between $d$-type and $p$-type states, which illustrates that the conversion between nontrivial and trivial bandgaps of the PC can be achieved by expanding and shrinking the R. It has been found that the relative sizes of $a_0$ and 3R tune the topological phase of the band structures [1,40]. When $a_0$=2.7R, the PC is nontrivial topology, while when $a_0$=3.52R, the PC is trivial topology. In this paper, only nontrivial topology is used to design, so we calculate the band structures for two cases with $a_0$=2.7R, 3.52R. The Brillouin zone is shown in the hexagonal inset in Fig. 1(a). When $a_0$=2.7R, the *d*-type states are at the Γ point of the lower band, and the *p*-type states are at the Γ point of the upper band, shown in Fig. 1(b). However, when $a_0$=3.5R, there is a band inversion, shown in Fig. 1(c).

In the existing works, the use of trivial-nontrivial [41] and trivial-nontrivial-trivial sandwich structures [42] of edge states for topological transport is the most common. Recently, topological waveguide transmission has been realized using a nontrivial PTIs-air-nontrivial PTIs in the microwave band [40] and a topological power divider has been realized using a nontrivial PTIs-air-nontrivial PTIs in the terahertz band [43], etc., which means that the introduction of line defects into nontrivial PTIs allows for efficient topological transmission. By using the nontrivial PC-gap-nontrivial PC structure, energy is transferred locally within the topological line defect region. The supercell of the proposed waveguide structure is shown in Fig. 2(a), which is symmetric about the plane mirror surface of *y*=0, and the blue rectangular boxed area indicates a periodic cell. Fig. 2(b) illustrates the calculated band structure by sweeping $k_x$ from -$\pi/a_0$ to $\pi/a_0$. We find that there are three bands within the band gap, which divide the band gap (21.502-25.236 GHz) into three parts. The first band has frequencies lower than 22.586 GHz, and the third band has frequencies higher than 22.938 GHz. Both the first and third bands have large group velocities near $k_x$=0, while the second band has nearly zero group velocity near $k_x$=0. And the width of line defect will have an effect on the band gap distribution, with the gap decrease, the third band (red curve) will move to higher frequencies and gradually disappear into the bulk, and with the gap increase, the first band (blue curve) will move to lower frequencies and gradually disappear into the bulk [40]. The closer the band structure is to the bulk, the more the topological edge state effects are relatively weakened. In addition, there is a tiny gap at the $k_x$=0 point in Fig. 2(b), which may result from the fact that for the expanded lattice case, the PC has a dimensional hierarchy of the higher-order topology in which both the first- and second-order photonic topological insulator emerge [44]. Fig. 2(c) gives the Poynting vectors at the six marked positions in Fig. 2(b), which are all mirror-symmetric about *y*=0, but different in the direction of energy flow. For the first and third bands, the direction of energy flow drive to the right along the adjacent supercell for the modes marked by triangles, and to the left along the adjacent supercell for the modes marked by inverted triangles. However, for the modes marked by a dot in second band, the energy flow forms a bound pattern inside the single cell in the counterclockwise direction on the left side of $k_x$=0 and in the clockwise direction on the right side of $k_x$=0. The first and third bands have similar modes to the helical edge states in the QSH effects, that is the direction of rotation correlated with the direction of energy flow. It implies pseudospin-locked unidirectional propagation. However, the second band of bound modes has no coupling path between adjacent cells, which might be useless for guided wave applications.

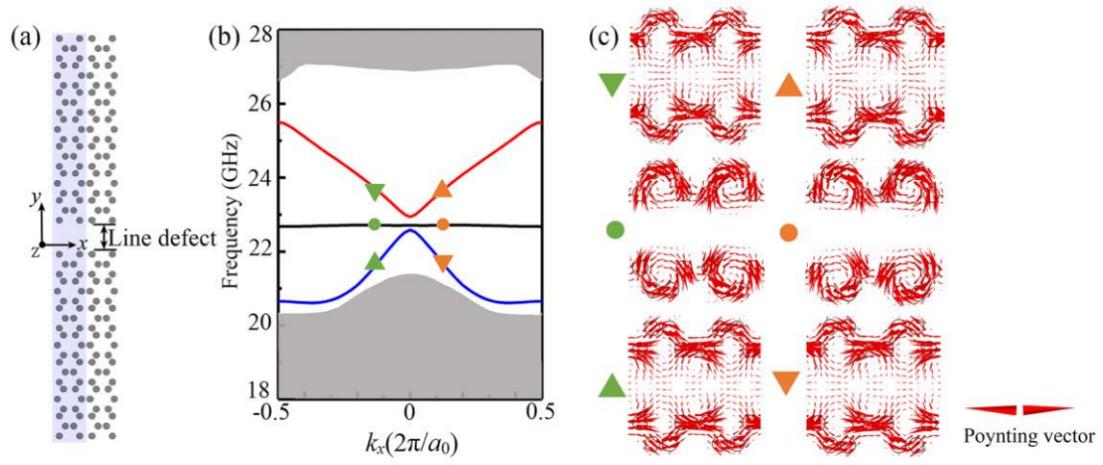

**Fig.2** Topological line defect states. (a) Supercell constructed for the proposed topological gap waveguide, and line defects are formed by removing a whole unit. The blue rectangular indicates the supercell structure of one periodic. (b) Corresponding band structure. (c) Poynting vector about the line defect at the marked points in (b).

### 3 Results and Discussion

To further clarify the robust transport of spin-locked topological line defect states, we designed two straight PC waveguides and a Z-type PC waveguide as shown in Fig. 3. The electric field distribution of three waveguides in $z =1.1h_0$ plane are shown in Figs. 3(a-c). The corresponding Poynting vectors are shown in Fig. 3(d-f), which shows that two nontrivial-PCs and the air gap (which can be regarded as a trivial-PC) form spin-locked topological edge states at the upper and down edges, meanwhile the EM field is also strong within the air gap channel, i.e. topological line defect state. The topological line defect state is robust to disorder and sharp bends of the structure. The two waveguides shown in Fig. 3(b-c) verify this property. Disorder is introduced by removing one of the cylinders at the upper edge of the gap as shown in Fig. 3(b), and a Z-type waveguide is designed with two 60° sharp bends in Fig. 3(c). The corresponding Poynting vectors are shown in Fig. 3(e-f). It can be seen that the Poynting vector distribution are distorted near the disorder and sharp bends, but reestablished after bypassing the defects. Evidently, the unidirectional transmission of EM waves is well maintained within the straight topological waveguides with presence of disorder and the Z-type waveguide. The transmission rates for the three cases are given in Fig. 3(g). The best transmission band for straight waveguide without disorder is 22.8-25.3 GHz, shown in the black curve in Fig.3(g). It can be seen that the transmission rate is basically unchanged for the straight waveguide with disorder compared to the straight waveguide without disorder. Compared to straight waveguide, Z-waveguide does not introduce more backscattering, but has a slightly narrower transmission bandwidth of 23.05-25.15 GHz. Moreover, the minimum transmission-loss of the topological gap waveguide reaches 0.12 dB. The above results demonstrate that the transmission of our designed topologically gap waveguide is robust to disorder and sharp bends of the structure.

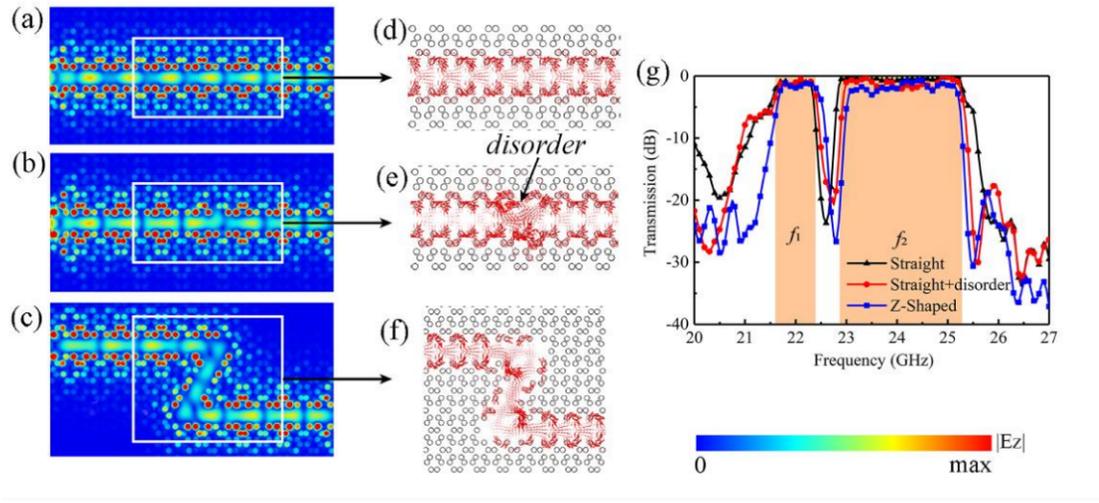

**Fig.3** Distribution of electric fields at the first/third band in Fig. 2(b), $f$ =24.35 GHz. (a) Straight waveguide. (b) Straight waveguide with disorder. (c) Z-shape waveguide. (d) (e) (f) Corresponding distribution of Poynting vector of (a), (b) and (c) respectively. (g) Transmission of three topological gap waveguides.

To further verify the backscattering immunity property of the proposed structure, we excited topological line defect states with a line source carrying orbital angular momentum (OAM). Two four-line source arrays with opposite modes are placed symmetrically in the cells at the top and bottom edges, as shown in the positions marked in Figs. 4(a-b). The topological line defect states of the first/third band in Fig. 2(b) can be selectively excited by setting the OAM with different symbols. The setting of the OAM in Fig. 4(a) is the same as the eigenstate shown in the triangle in Fig. 2(c), thus exciting the spin-locked topological line defect states propagating to the right. While the setting of the OAM in Fig. 4(b) is the same as the eigenstate marked by the inverted triangle in Fig. 2(c), thus exciting the spin-locked topological line defect states propagating to the left. Figs.4(a-b) give the distribution of electric fields inside the waveguide at $f$ =24.35 GHz. The corresponding energy flow direction is shown in Figs. 4(c-d), which shows that the proposed structure is immune to backscattering. The above study shows that the gap waveguide structure we have designed is indeed spin-locked for unidirectional transmission, robustness, and better transmission efficiency.

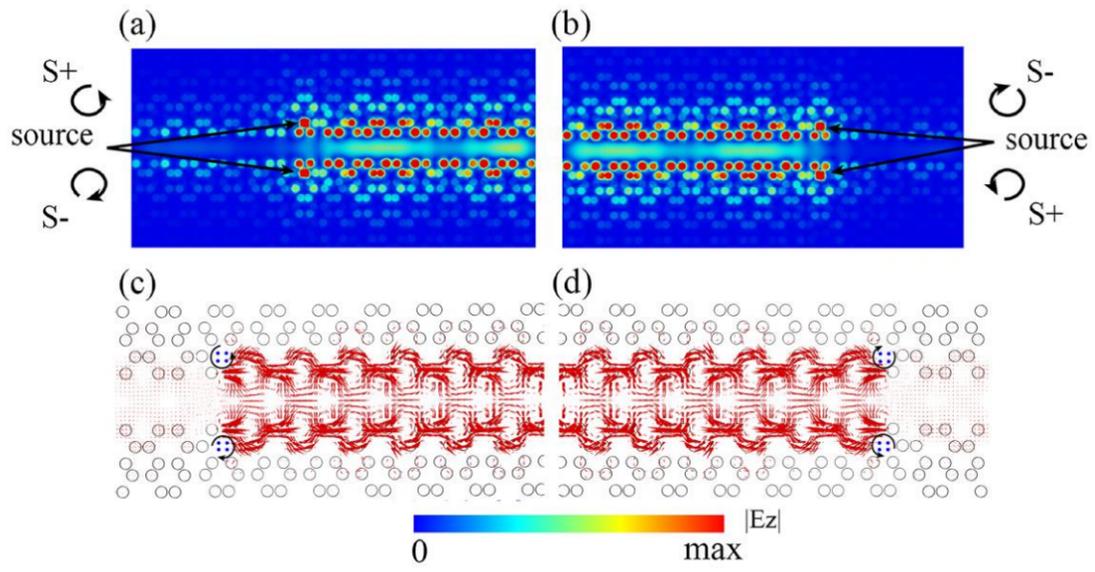

**Fig.4** Distribution of electric fields at the first/third band in fig.2(b), $f$ =24.35 GHz. (a) A pair of four-line source array carrying reversed OAM, which are located at near the edge of the hexagon, and the OAM setting of the source is the same as the eigenstates marked by the triangle in Fig. 2(b). (b) The OAM setting of the source is the same as the eigenstates marked by the inverted triangle in Fig. 2(b). (c)(d) Corresponding Poynting vector distribution respectively.

In addition, to reconfigure our design dynamically on-demand, we implement a tunable topological gap waveguide using the adjustable anisotropy of the nematic LC, shown in Fig. 5(a). Due to a gap left between the top of peg-like topological pattern and the top plate, they are direct current decoupled. The effect of metal plates is to provide two electrodes of external circuits and LCs are to provide two states of *off* or *on*, i.e., similar to a switch, shown in Figs. 5(b-c). Thus, the top and bottom plates can be used as electrodes for vertical alignment of the LC, and using the anchoring forces on the surfaces of the top and bottom plates to achieve horizontal alignment for LC [36,45]. The macroscopic relative permittivity $\varepsilon_r$ and loss tangent tan $\delta$ can be changed by controlling the orientation of the LC molecules with respect to the applied radio frequency field. The relationship between the relative permittivity $\varepsilon_r$ and the inclination angle of the LC molecular orientation (the angle between the long molecular axis and the axis of the direction of the applied RF field, i.e., the angle between the long molecular axis and the *z*-axis) is given by [36]:

$$\varepsilon_r = \begin{bmatrix} \varepsilon_{r,\parallel}\sin^2(\alpha) + \varepsilon_{r,\perp}\cos^2(\alpha) & 0 & \frac{1}{2}(\varepsilon_{r,\parallel} - \varepsilon_{r,\perp})\sin(2\alpha) \\ 0 & \varepsilon_{r,\perp} & 0 \\ \frac{1}{2}(\varepsilon_{r,\parallel} - \varepsilon_{r,\perp})\sin(2\alpha) & 0 & \varepsilon_{r,\parallel}\cos^2(\alpha) + \varepsilon_{r,\perp}\sin^2(\alpha) \end{bmatrix} \quad (2)$$

where α is the angle between the long molecular axis and the *z*-axis, and the permittivity $\varepsilon_{r,\parallel}$, $\varepsilon_{r,\perp}$ (with respect to the long molecular axis) are the permittivities in the two extreme cases of parallel and orthogonal RF polarization, respectively, determined by the material itself. When α=90°, the LC molecules are arranged oriented horizontally:

$$\varepsilon_r = \begin{bmatrix} \varepsilon_{r,\parallel} & 0 & 0 \\ 0 & \varepsilon_{r,\perp} & 0 \\ 0 & 0 & \varepsilon_{r,\perp} \end{bmatrix} \quad (3)$$

And when α=0°, the LC molecules are aligned in the vertical direction:

$$\varepsilon_r = \begin{bmatrix} \varepsilon_{r,\perp} & 0 & 0 \\ 0 & \varepsilon_{r,\perp} & 0 \\ 0 & 0 & \varepsilon_{r,\parallel} \end{bmatrix} \quad (4)$$

For microwave applications requiring a combination of lower loss and higher dielectric anisotropy, the GT series of LC mixtures are suitable for research, where the relative permittivity of GT7-29001 in the two extreme cases are $\varepsilon_{r,\parallel}$=3.53, $\varepsilon_{r,\perp}$=2.46, respectively. Therefore, GT7-29001 was chosen for the study of this work, because it has large anisotropy, fast response time, and still low loss [37]. We plot the band structure of supercell when the external circuit is arranged from *off* to *on* during several cases of special inclination of LC molecules, as shown in Fig. 5(d). It is clear that as the angle decreases, the transmission band of the structure moves towards lower frequencies. The transmission bands possess completely staggered frequency bands when the $\alpha = 90°\ or\ 0°$.

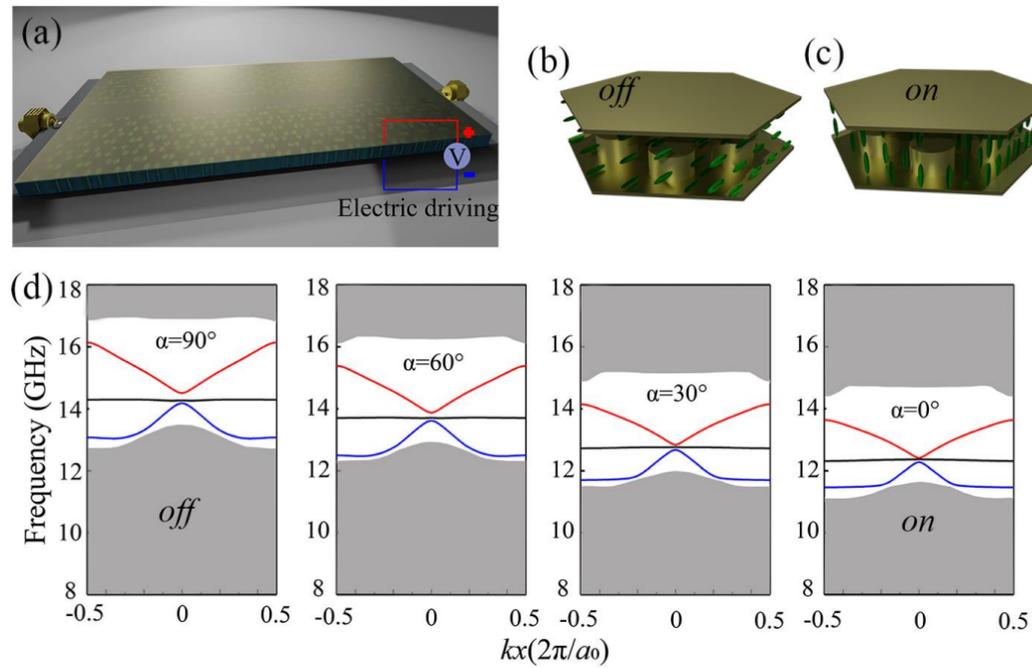

**Fig.5** Schematic diagram of a tunable topological gap waveguide structure. (a) Structure is composed of a triangular lattice array. The metallic cylinders (yellow) are wrapped by LCs (blue) and covered by the metal plates (yellow) both top and bottom. The metal plates are used as external circuit electrodes for driving the phase change of the liquid crystal. (b) Arrangement of LC molecules when the external circuit is in the *off* state. (c) Arrangement of LC molecules when the external circuit is in the *on* state. (d) Corresponding band structure, the LC parameters for the four different cases are given by equations (2-4).

In order to explore the transmission performance of this waveguide after filling the LC, three different waveguides were also designed. Their electric field distribution, energy flow, and transmission efficiency are analyzed in terms of their differences from the structure without the filled LC. The corresponding simulation results are given in Figs. 6 (a-g). Comparing the results of Figs. 6(a-f) with those of Figs. 3(a-f), it clearly follows that the transmission characteristics of the gap waveguide with and without the LC case are largely maintained. However, comparing the results of Fig. 6(g) with those of Fig. 3(g), it is evident that some loss is introduced and the bandwidth is narrowed after filling the LC. Based on the above study, after filling LC as the background material, although the transmission bandwidth is slightly reduced and a certain insertion loss is introduced, it shifts the transmission band to lower frequencies, which has a greater prospect of application in the miniaturization of devices.

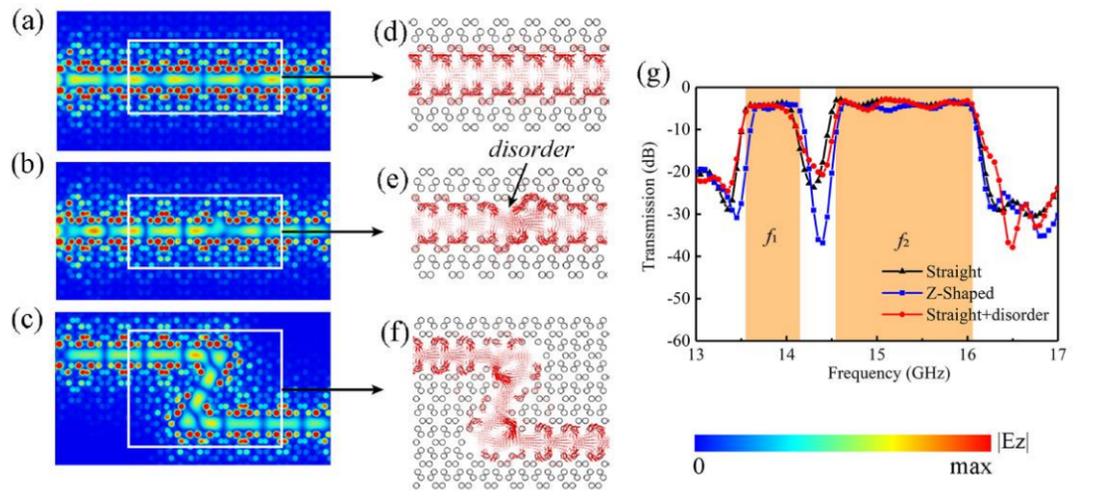

**Fig.6** Distribution of electric fields at the first/third band in Fig. 5(d), $f$ =15.2 GHz. (a) Straight waveguide. (b) Straight waveguide with disorder. (c) Z-shape waveguide. (d) (e) (f) Corresponding distribution of Poynting vector of (a), (b) and (c) respectively. (g) Transmission of three topological gap waveguides.

Finally, to further explore the regulation mechanism of LC molecules when the external circuit is in different states. As shown in Fig. 7(a), the variation of the band structure of the PC with the inclination angle α of the LC molecular arrangement orientation is given. It can be seen that α varies from 90° to 0° as the external circuit transitions from the *off* to the *on* state, and the process is reversible. As the α angle decreases, the band gap shifts to lower frequencies while the bandwidth becomes slightly narrower, but its band characteristics do not change. It is the change in the relative permittivity $\varepsilon_r$ and loss tangent tan $\delta$ of the LC due to the transition from the *off* to the *on* state of the external circuit that leads to this result. Based on the above discussion, we can design reconfigurable PC waveguides using reversible changes in the orientation arrangement of nematic LC molecules. We design a straight waveguide and give its transmission characteristics at several special inclination angles, as shown in Fig. 7(b). The results are in agreement with the results of the band structure characteristics corresponding to the Fig. 7(a). Figs. 7(c-d) show the distribution of the electric field inside the waveguide at $f$ =15.2 GHz when the external circuit is in the *off* or *on* states, respectively. Obviously, when the external circuit is in the *off* state, EM waves with a frequency of 15.2 GHz are transmitted robustly within the waveguide, but when the external circuit is in the *on* state, it cannot be transmitted within the waveguide. Therefore, the topological line defect states of the designed gap waveguide can be turned *on* and *off* by controlling the external circuit, thus controlling the opening and closing of the frequency-specific electromagnetic wave transmission path for reconfigurability.

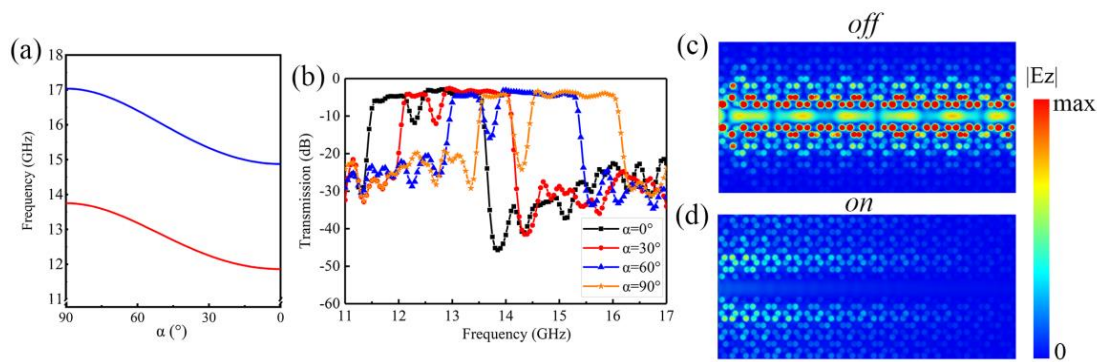

**Fig.7** (a) Variation of the unit-cell band structure of a PC with the inclination angle of the LC molecular orientation, (b) Variation of transmission rate of straight waveguide with α, (c) the distribution of the electric field inside the straight waveguide at *f* =15.2 GHz, when the external circuit is in the *off* state, (d) the distribution of the electric field inside the straight waveguide at *f* =15.2 GHz, when the external circuit is in the *on* state.

**4 Conclusion**

In summary, a topological gap waveguide based on spin-unidirectional locking was designed using the non-trivial states of PTIs. The introduction of line defects into the gap waveguide structure constructs topological line defect states similar to the topological edge states formed by trivial and nontrivial photonic crystals, which are immune to backscattering, unidirectional locking of energy transfer, and robust to sharp bends and disorder. The waveguide combines the advantages of both topological and gap waveguides, and it not only has efficient and robust transmission, but also achieves a minimum transmission-loss of 0.12 dB. In addition, a reconfigurable topological gap waveguide has been realized using the anisotropic tunability of the nematic LC, along with a switch between open and closed states of the transmission channel. The proposed waveguide structure can be implemented using conventional semiconductor technology and can be used to integrate photonic devices, topological optical circuits, and optical communication systems. Furthermore, it has potential applications in achieving CMOS compatibility and backscattering suppression in tunable on-chip integrated topological photonic devices.

**Acknowledgements**：All the authors acknowledge financial supports from Fundamental Research Funds for the Central Universities (CCNU19TS073, CCNU20GF004); Hubei Province Technology Innovation Plan Key Research and Development Project; Young Scientists Fund (NSFC11804087); Science and Technology Program of Hubei Province (2022CFB553, 2022CFA012, 2018CFB148); Hubei Provincial Department of Education (T2020001); and Hubei University (030-017643).

**Data Availability:** The data that support the findings of this study are available from the corresponding author upon reasonable request.